\documentclass[a4paper,12pt]{article}
\pdfoutput=1 

\usepackage{mathrsfs,graphicx,rotating,amsmath,amsfonts,mathtools,booktabs,amssymb,wasysym,caption}
\usepackage{hyperref}
\usepackage{slashed}
\usepackage[table,xcdraw,dvipsnames]{xcolor}
\usepackage{graphicx}
\usepackage{bbold}
\usepackage[utf8x]{inputenc}
\usepackage[english]{babel}
\usepackage{multirow,multicol}
\usepackage{epstopdf}
\usepackage{bbm}
\usepackage{changepage}
\usepackage{appendix}
\usepackage{libertine}
\usepackage{braket}
\usepackage{wasysym}
\usepackage{empheq}
\usepackage{cancel}
\usepackage{enumitem}%
\usepackage{mathrsfs}
\usepackage{lmodern}
\usepackage{tabularx}
\usepackage{multicol}
\usepackage{color}
\usepackage{mathtools}
\usepackage{verbatim}
\usepackage{arydshln}
\usepackage{url}
\usepackage[normalem]{ulem}
\usepackage{youngtab}

\begin{document} 


\begin{center}
\vspace{1cm}
{\Large \bf A view of flavour physics  in 2021} \\
\vspace{2cm}
{\large Riccardo Barbieri} \\

{\it \small Scuola Normale Superiore, Piazza dei Cavalieri 7, 56126 Pisa, Italy} \\
\vspace{1cm}
\end{center}
Based on a view of current flavour physics and motivated by the hierarchy problem and by the pattern of quark masses and mixings, I describe  a picture of flavour physics that should give rise in a not too distant future to observable deviations from the SM in Higgs compositeness and/or in  B-decays with violations of Lepton Flavour Universality, as hinted by current data, or perhaps even in supersymmetry, depending on the specific realisation. 
\\
\vspace{4cm}
\begin{center}
A contribution to:\\
The special volume of Acta Physica Polonica B to commemorate \\
 Martinus Veltman
\end{center}

\newpage

\section{A master}

I  met Martinus Veltman  for the first time in 1970  when, together with Ettore Remiddi,  we went to visit him in Utrecht. We were looking for help to solve some problems we had in the use of Schoonschip to perform analytic calculations of higher order QED processes. Two years later Veltman came to the SNS in Pisa to give lectures on field theory, which I attended, that became the basis for the CERN yellow report in 1973  under the  title of Diagrammar with Gerard 't Hooft as coauthor. All this is recalled by Ettore in an article also to appear on this volume, with particular reference to the various tools - Schoonschip, dimensional regularisation, the largest time equation - introduced by Veltman and crucial to our work throughout the seventies in QED and QCD. Since then I have met Veltman many other times, especially, but not only, at CERN. 

As far as I can tell, the subject of flavour, considered here, has not been particularly important in all the work of Veltman, except for a few early papers that he wrote in the sixties. Given the key  pioneering role that Veltman played in the ElectroWeak precision tests, however, one may find a connection with his work in the fact that I underline the importance of precision measurements in flavour. True enough, Veltman always liked precise calculations, whereas here I am often bound to semi-quantitative considerations based on a "picture" rather than on a defined theory. Based on this picture,  however, I argue that the experimental flavour program in the next decade or so might help to design a complete theory of flavour, thus allowing for precise calculations in this area as well.

\section{Introduction and motivations}
\label{Intro}

To the extent that the level of my perception has been constant in time, I feel pretty sure to say that this is the relatively most uncertain time that I have seen in the last fifty years of Particle Physics (PP). To a large extent, although not exclusively, this is due to the contrast between the  impressive and steadily increasing experimental success of the Standard Model (SM) since its complete formulation in the early seventies and, on the opposite side, the yet unsolved "structural" problems of the SM itself. The sharpest of these problems, in my view, is the unpredictability of the masses of all its massive particles, 15=17-2 (although with a  number of  independent observables correlated by $m_W, m_Z$, and also, via loops, by $m_t$ and $m_H$). Among them there is the Higgs mass, $m_H$, with the specific property of being sensitive to any UV  scale of physics to which the Higgs boson couples: the "hierarchy problem". 

Where  could some light come from? Perhaps a theoretical breakthrough in Beyond the SM - not that one hasn't already tried, sometimes with great ideas, none of which, however, confirmed by experiments so far - or  in the very foundations of the SM, like Field Theory or even Quantum Mechanics.  A possible new finding in astrophysics or cosmology with some connection with PP. In this area the fundamental nature of issues like  Dark Matter or the origin of the Baryon Asymmetry in the Universe is not disputable. The open question, however, is whether or not their solution will have an impact on the structural problems of the SM.
Or, finally, an experimental deviation from the  SM in the form of a new particle or of a  discrepancy with  the SM in some precision Low Energy observables.

\begin{figure}[t]
\centering
\includegraphics[clip,width=.85\textwidth]{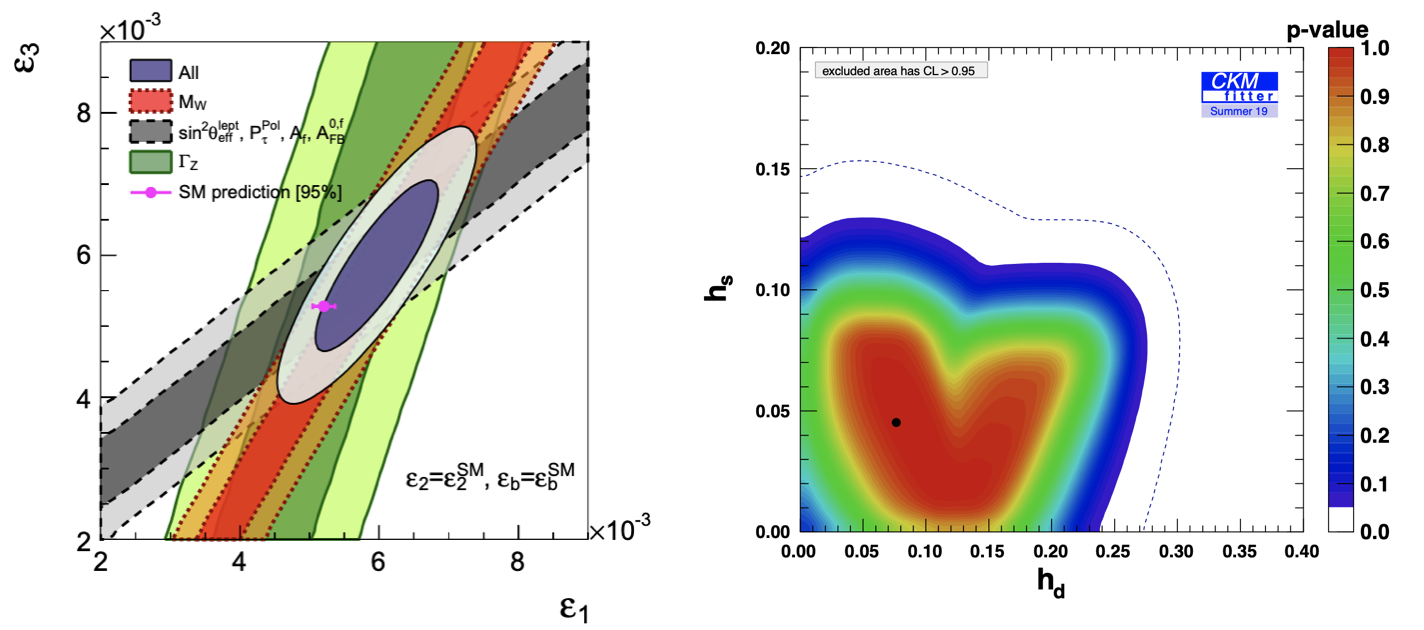}
\caption{Left: The ElectroWeak observable parameters $\epsilon_{1,3}$, taken from Ref.~\cite{deBlas:2019okz}. Right: The relative deviations from the SM in $M_{12}(B_d)$, $h_d$, and $M_{12}(B_s)$, $h_s$, taken from Ref.~\cite{Charles:2020dfl}.
}
\label{fig:eps_DeltaB=2}
\end{figure}

The uncertainty of the current situation motivates the widest possible approach. This is indeed what is happening, as far as I can tell. Here, with a time horizon of a decade or so or, most likely, before the operation of the next High Energy collider, I focus my attention to the potential of the precision programme in flavour physics, based on a picture of the origin of charged fermion masses that I shall describe.
  The underlying assumption is that there is new physics in the MultiTeV
energy region. The motivation for such an assumption is both generally exploratory and suggested by the still pending hierarchy problem, mentioned above. Depending on the specific realisation of the flavour picture, observable deviations from the SM might emerge in Higgs compositeness and/or in  B-decays with violations of Lepton Flavour Universality, as hinted by current data, or perhaps even in supersymmetry.

As a specific motivation of this programme I find useful the two Figures~\ref{fig:eps_DeltaB=2}, taken   from Refs.~\cite{deBlas:2019okz,Charles:2020dfl}, where one can compare the current status of some key precision observables in ElectroWeak Physics, the parameters $\epsilon_{1,3}$~\cite{Altarelli:1990zd,Altarelli:1991fk}, and of the relative deviations from the SM, $h_d$ and $h_s$, in $\Delta B_d=2$ and $\Delta B_s=2$ transitions. As one sees, the current relative precision, respectively sensitive to "ElectroWeak loops" and to "Flavour Changing Neutral Current loops", is in both cases at the level of $10\div 20\%$. Taking into account experimental, theoretical and parametric uncertainties, at a future facility like FCC-ee the  foreseen relative precision on $\epsilon_{1,3}$ might reach the $2\div 4\%$ level~\cite{deBlas:2019rxi}. A similar level of relative precision could be reached on $\Delta B_d=2$ and $\Delta B_s=2$ by an intense flavour physics programme in the next 10-15 years, as shown in \cite{Charles:2020dfl}.

\begin{figure}[t]
\centering
\includegraphics[clip,width=.40\textwidth]{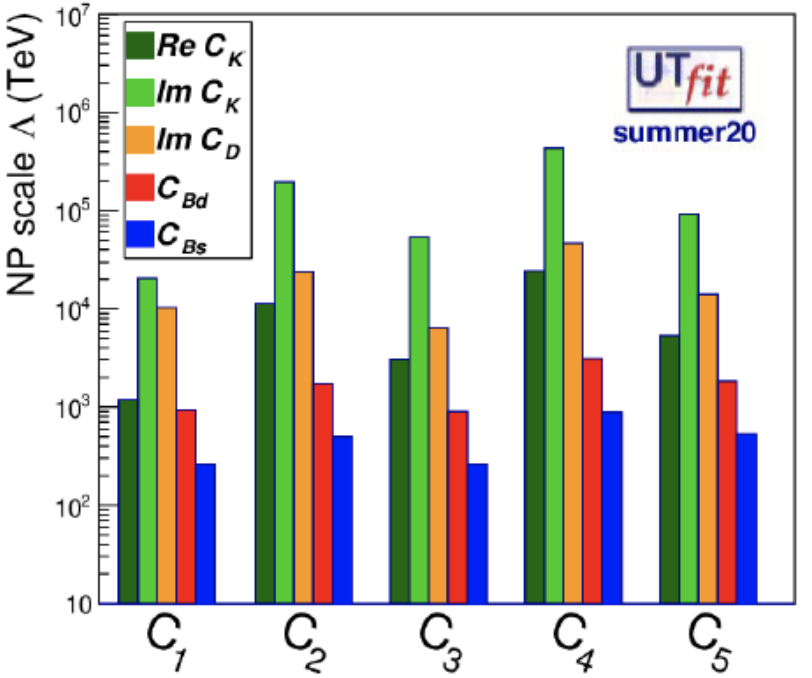}
\caption{Lower bounds on the scale associated with 4-fermion, $\Delta F=2$ operators in different Lorenz configurations, as defined in~\cite{Gabbiani:1996hi}: $\mathcal{O}_1=(\bar{q}_{Li}\gamma_\mu q_{Lj})^2$;  $\mathcal{O}_{2,3,4,5}$ contain admixtures of left and right fermion fields. Courtesy of Luca Silvestrini.
}
\label{fig:DeltaF=2}
\end{figure}

An apparent obstacle to the programme just outlined seems well illustrated in Fig.~\ref{fig:DeltaF=2}, where one shows the lower bound on the scales associated with 4-fermion, dimension-6 operators which violate flavour by two units in different Lorenz configurations.
The implications of these bounds are apparently clear. Flavour is determined by very high energy physics; any new MultiTeV physics, if existent at all, is flavour independent; the hierarchy problem is not related to flavour physics. The flavour picture outlined in the next Section provides an alternative view.
Broadly speaking it is inspired by old ideas, like Extended Technicolour, or less old ones like Extra-dimensions with branes at different positions~\cite{Dvali:2000ha}, or by more recents attempts, as, e.g., in Ref.s~\cite{Panico:2016ull,Bordone:2017bld}.

\section{A 3 Scale Flavour Picture }
\label{AFP}

\begin{figure}[t]
\centering
\includegraphics[clip,width=.38\textwidth]{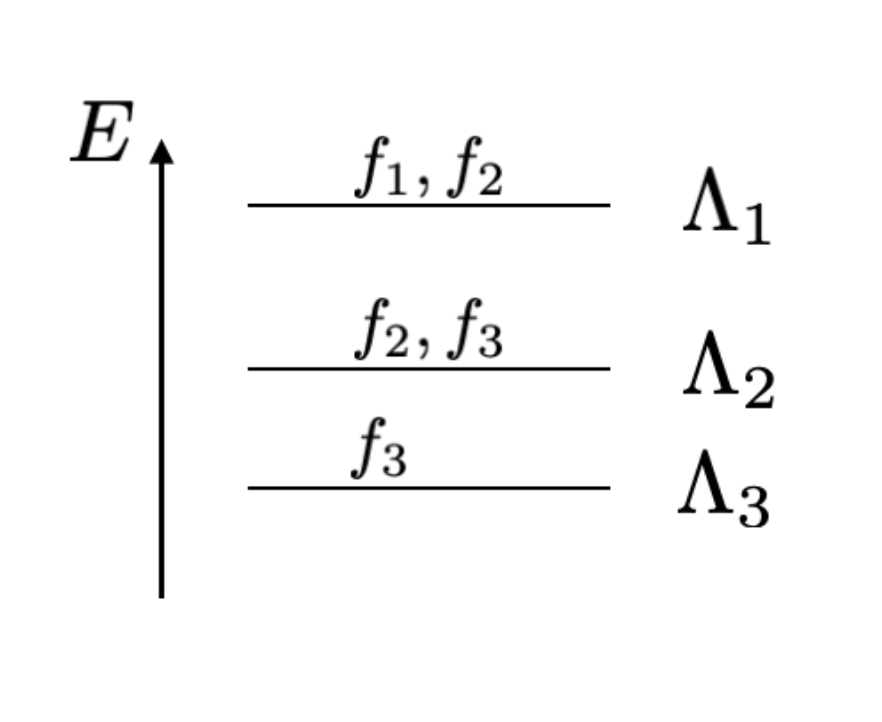}
\caption{A scheme of the 3 Scale Flavour Picture as described in the text
}
\label{fig:FP}
\end{figure}

Based on the hierarchies  of charged fermion masses and quark mixing angles, I assume that the three different families are distinguished by new interactions at (at least) three different scales, $\Lambda_3<\Lambda_2<\Lambda_1$, from where  their masses originate.
Denoting collectively by $f_i$ the $SU_{3,2,1}$ fermion multiplets of the i-th generation, $f_i= (q,u,d,l,e)_i$ in standard notation, the pattern of masses and mixings is supposed to arise, at least in part, from the fact that at $\Lambda_i$ no fermion  $f_j$ is involved with $j<i$. This is summarised in Fig.~\ref{fig:FP}, where I also assume "nearest-neighbour" interactions, as implied, e.g., by the existence of an accidental discrete symmetry, $f_3\rightarrow - f_3$ at $\Lambda_1$.
The presence of more than one family at $\Lambda_{2,1}$ is what gives rise to mixings. I ignore two issues: the hierarchy of masses inside one generation, which may be due to the existence of more than three scales, and the origin of the mixing in the lepton sector, which is far from hierarchical and may be attributed to the scale(s) and to the peculiar structure of the right-handed neutrino mass matrix, of different origin than the mass matrices of the charged fermions.

The dominant manifestations at low energy, $E< \Lambda_3$, of this picture will be:
\begin{itemize}
\item The effects of higher dimensional operators scaled by $\Lambda_3$ and only involving the third generation in the interaction basis;
\item An accidental $U(2)^n$ symmetry, under which the $f_3$ are singlets and $f_{1,2}$ transform as doublets~\cite{Barbieri:2011ci}, progressively broken at $\Lambda_{2,1}$, which controls the Yukawa couplings and the consequent involvement of the first two generations, in the mass basis, at the scale $\Lambda_3$.
\end{itemize}

I expect that specific realisations of this flavour picture will be relevant in presence of other new physics in the MultiTeV with consequently different specific manifestations. In the following Sections I consider in turn the cases of Composite Higgs, of B-decays with Lepton Flavour Violation (if  confirmed by data), which may in fact be coupled with Higgs compositeness itself, or even the case of supersymmetry.
A central question in all of these cases is how low can the scale $\Lambda_3$ be consistently with current constraints. Unless otherwise stated, I shall ignore possible low energy effects of the interactions at $\Lambda_{2,1}$ (except for their contribution to the Yukawa couplings) assuming that they do not lead to more stringent  constraints than the interactions at $\Lambda_3$ itself.

\subsection{Remaining uncertainties}
\label{RU}

To be able to determine  in a  concrete way the bounds on the scale $\Lambda_3$ in the different situations, one needs the diagonalisation matrices of the different Yukawa couplings
\begin{equation}
Y^U = (U^L)^+Y^U_{diag}U^R,\quad
Y^D = (D^L)^+Y^D_{diag}D^R,\quad
Y^E = (E^L)^+Y^E_{diag}E^R.
\label{diag_m}
\end{equation}
The picture of Fig.~\ref{fig:FP} implies a hierarchical structure of $Y^{U,D,E}$ and the vanishing of their $Y_{13}, Y_{31}$ elements. In turn, up to small relative errors controlled by the precise knowledge of $Y^{U,D,E}_{diag}$, every diagonalisation matrix in eq.~(\ref{diag_m}), collectively called $\mathcal{U}$, reduces to the product of two subsequent unitary transformations in the 12 and 23 sectors, $\mathcal{U}=\mathcal{U}(12)\mathcal{U}(23)$. 

Taking advantage of $V_{CKM}=U^L(D^L)^+$, on can show~\cite{Barbieri:2011ci} that $U^L$ and $D^L$ are determined up to one physical phase and  their 23 elements,  constrained by
\begin{equation}
V_{cb}=U_{23}^L-D_{23}^L.
\end{equation}
This last ambiguity can be fixed by invoking an accidental discrete symmetry at $\Lambda_2$, $d_3\rightarrow - d_3$ or $u_3\rightarrow - u_3$, giving respectively $D_{23}^L=0$ or $U_{23}^L=0$.

The right handed diagonalisation matrices, in particular $U^R, D^R$, remain unknown. They can leave  the third generation  almost fully decoupled from the first two, i.e. $\mathcal{U}^R(23)\approx \mathcal{D}^R(23)\approx \bold{1}$, by demanding, again at $\Lambda_2$, that $q_3\rightarrow - q_3$ be an invariance. 
All of this is effectively equivalent, in the quark sector, to a "minimally violated $U(2)^3$" flavour symmetry~\cite{Barbieri:2011ci}.

\section{Flavour in Composite Higgs}
\label{SCH}

In this Section I assume, as customary in the literature,  that a new strong interaction with a confinement scale $m_*$ and a strong coupling $1< g_* <4\pi$ gives rise, after spontaneous symmetry breaking, to the Higgs, $H$, as a Pseudo-Nambu-Goldstone-Boson (PNGB)~\cite{Agashe:2004rs}.  The standard fermions do not feel directly this new strong interaction, but, to get a mass, they are connected at the scales $\Lambda_{3,2,1}$, as specified above,  with a composite operator of the strong sector, $\mathcal{O}_H$, with $<0|\mathcal{O}_H|H>\neq 0$. The strongest of these couplings occurs at $\Lambda_3$ 
\begin{equation}
\mathcal{L}^{top}_Y(\Lambda_3)=\frac{x_L x_R}{\Lambda_3^{d_H-1}}  \bar{q}_3 \mathcal{O}_H u_3,
\end{equation}
and gives rise to the top Yukawa coupling $y_t$ at the compositeness scale $m_*$, after $\mathcal{O}_H\rightarrow g_* m_*^{d_H-1} H$,
\begin{equation}
y_t=g_* x_L x_R (\frac{m_*}{\Lambda_3})^{d_H-1}.
\label{topY}
\end{equation}
Here $d_H$ is the anomalous dimension of the operator $\mathcal{O}_H$ and $x_{L,R}\leq 1$ are dimensionless parameters, universally associated with $q_3, u_3$, which measure their respective "level of compositeness". 
Since $y_t\approx 1$ and $d_H\lesssim 2$~\cite{Rattazzi:2008pe}, eq.~(\ref{topY}) suggests that $\Lambda_3$ be identified with the compositeness scale $m_*$. The question to ask is which are the main effects in flavour physics that one expects from this picture at low energy with respect to $\Lambda_3\approx m^*$.

 \begin{table}[t]
 \small
$$\begin{array}{c|c|c|c|c}
Observable&Operator&\Lambda/TeV (pres) &\Lambda/TeV  (fut)&  m_*/\Lambda\\ \hline
\epsilon_K&\mathcal{Q}_1^{sd} = (\bar{s}_L\gamma_\mu d_L)^2 &7&13&x_t\\ \hline
\Delta M_{B_d}&\mathcal{Q}_1^{bd} = (\bar{b}_L\gamma_\mu d_L)^2 &  8&18&x_t\\ \hline
\Delta M_{B_s}&\mathcal{Q}_1^{bs} = (\bar{b}_L\gamma_\mu s_L)^2 & 9&20&x_t\\ \hline
\Delta M_D, p/q &\mathcal{Q}_1^{cu} = (\bar{c}_L\gamma_\mu u_L)^2 & 3&10 &x_t\\ \hline \hline
b\rightarrow s ~\bar{l}l& (\bar{s}_L\gamma_\mu b_L) H^+ i \mathcal{D}_\mu H   & 4.5&12&\sqrt{g^*x_t} \\ \hline
s\rightarrow d ~\bar{l}l(\bar{q}q)  & (\bar{d}_L\gamma_\mu s_L) H^+ i \mathcal{D}_\mu H   &1.7&5& \sqrt{g^*x_t} \\ \hline \hline
neutron EDM ~( *)&\approx m_t (\bar{t}_L\sigma_{\mu\nu}T^a t_R)g_SG^{\mu\nu}_a &\approx 5.5&16& g^*/4\pi\\ \hline \hline
electron EDM ~( *)&\approx m_t (\bar{t}_L\sigma_{\mu\nu} t_R) eF^{\mu\nu} & \approx 50 &? & \sqrt{g^*x_t} /(4\pi)\\ \hline \hline
\end{array}$$
\caption{Estimated lower bounds, present and future, on the scale associated with representative operators, from different observables. See text.
}
\label{Tab:CH}
\end{table}

As well known and manifest from Fig.~\ref{fig:DeltaF=2} , the strongest constraint from $\Delta F=2$ transitions without a flavour structure comes from $
\Delta S=2$,  4-fermion, LR interactions. To keep this constraint under control, I assume the symmetry $q_3\rightarrow - q_3$  at $\Lambda_2$, which decouples in particular $d_3$ from $d_{1,2}$, but also $u_3$ from $u_{1,2}$.  This leaves, upon use of eq.~(\ref{topY}), 
\begin{equation}
\mathcal{L}^{(4f)}  =\frac{y_t^2 x_t^2}{m_*^2} 
 (\bar{q}_{3}\gamma_\mu  q_{3})^2, \quad\quad
 \frac{y_t}{g^*} < x_t=\frac{x_L}{x_R}<\frac{g^*}{y_t}
 \label{4f}
 \end{equation}
 as the only possible source of $\Delta F=2$ transitions in the mass basis, controlled by the mixing matrices $U^L$ and $D^L$\footnote{For the purpose of the estimates of interest here, I neglect a possible cancellation between the operator in eq.~(\ref{4f}) and $(\bar{q}_{3}\gamma_\mu \sigma_a q_{3})^2$.}. 
 
 To estimate the bounds on $\Lambda_3\approx m^*$ from the various $\Delta F=2$ transitions is now a straightforward matter, starting from the bounds in Fig.~\ref{fig:DeltaF=2} and using $U^L$ and $D^L$ as determined in the previous Section. At least for concreteness, I use $U^L_{23}=0$ or $D^L_{23}=0$ when considering the bounds from $\Delta S=2,1$ and $ \Delta B=2,1$ or $\Delta C=2$ respectively. As recalled in Section~\ref{RU}, this leaves one single phase free in $U^L$ and $D^L$, which however does not affect the $\Delta S=2$ nor the $\Delta C=2$ transitions, where the distinction between CP-conserving and CP-violating effects is most crucial. 
 
 \begin{figure}[t]
\centering
\includegraphics[clip,width=.98\textwidth]{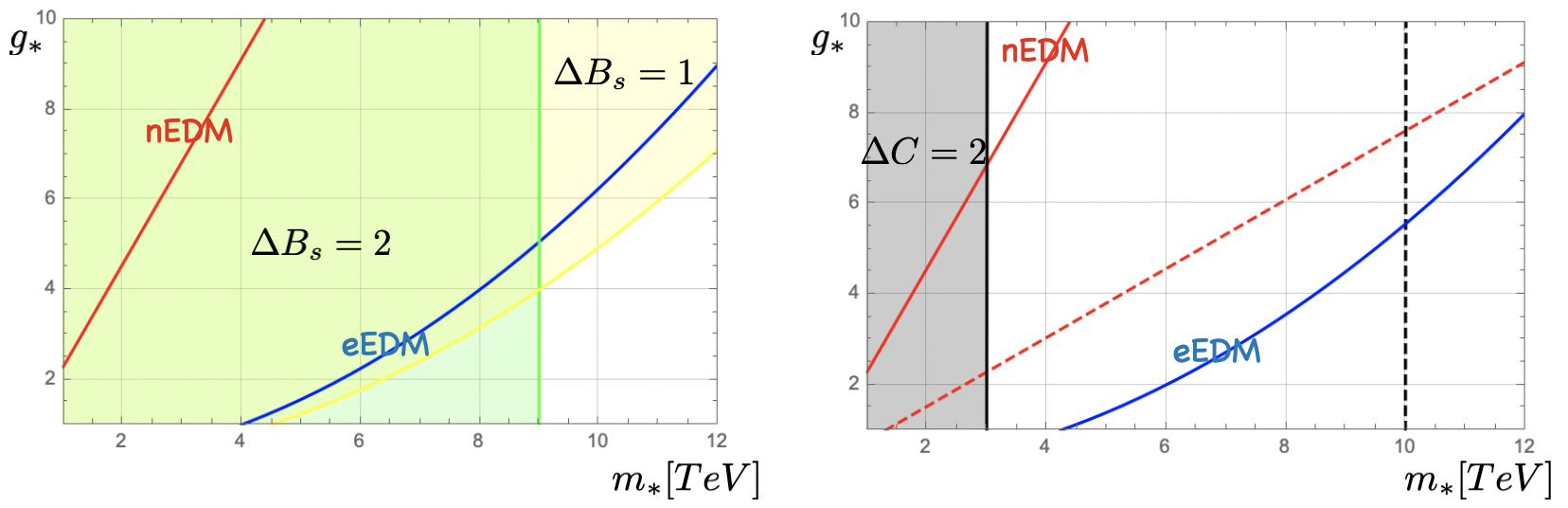}
\caption{Left: Estimate of the currently excluded regions for $U^L_{23}=0$. Right: Estimate of the currently excluded regions and of the foreseen sensitivity (dotted lines)  for $D^L_{23}=0$. The lines associated with the electron and the neutron EDMs assume maximal phases. Everywhere $x_t=1$.
}
\label{fig:Higgs_comp}
\end{figure}
 
 Along similar lines one can extract bounds on $\Lambda_3\approx m^*$ from $\Delta F=1$ transitions and from the neutron and electron Electric Dipole Moments (EDM),  although with an increasing level of uncertainty and model dependence both in $\Delta F=1$ transitions and even more so in the EDMs.
 Table~\ref{Tab:CH} summarises an estimate of these bounds, including a future projection,  from different observables. In the third and fourth columns, denoted $\Lambda/TeV$, I set $g^*=x_t=1$ and I ignore a strong loop factor $(g^*/4\pi)^2$ in the EMDs. The final column gives the relation between $\Lambda$, thus defined, and  $m^*\approx\Lambda_3$.
 An overall illustration of these bounds is given in Fig.~\ref{fig:Higgs_comp}, both for the case $U^L_{23}=0$ or $D^L_{23}=0$\footnote{The bound from CP violation in $\Delta C=2$ does not yet include the recent LHCb results~\cite{Pajero}}.
 
In this Section I have neglected direct significant couplings of the leptons to the composite sector, i.e. to the scale $\Lambda_3$; an assumption to be changed in the following Section.

\section{Anomalies in B-decays}
\label{AiBd}

If the third generation leptons are  involved as well with a significant strength in the new interactions at $\Lambda_3$, violations of Lepton Flavour Universality become of clear interest, in particular due to the theoretical cleanness of observables sensitive to them. This unavoidably brings the attention to the persistent, though not yet decisive, observations of anomalies  in semi-leptonic B-decays, especially since, at first sight, they occur with a pattern that looks quite compatible with the picture described above: 
\begin{itemize}
\item A roughly $10\%$ deviation from a tree level SM process in the charged channel $b\rightarrow c~l\nu$, quantified in the ratio~\cite{Abdesselam:2019dgh,Amhis:2019ckw,Bernlochner:2021vlv}
\begin{equation}
R_{D^{(*)}}=\frac{BR(B\rightarrow D^{(*)}\tau\nu)}
{BR(B\rightarrow D^{(*)}l\nu, l=\mu,e)},
\end{equation}
where, in the numerator, only one second generation particle is involved (and three in the denominator, neglecting corrections to the electron component);
\item A roughly $10\%$ deviation from a loop induced SM process in the neutral channel $b\rightarrow s~ll$, quantified in the ratio~\cite{Aaij:2021vac,Aaij:2017vbb}
\begin{equation}
R_{K^{(*)}}=\frac{BR(B\rightarrow K^{(*)}\mu\mu)}
{BR(B\rightarrow K^{(*)} ee)},
\end{equation}
where, in the numerator, three second generation particles are involved. In fact the anomaly in $R_{K^{(*)}}$ appears consistently reinforced by the measurements of the angular distributions in $B\rightarrow K^*\mu\mu$ and of $BR(B_s\rightarrow \mu\mu)$~\cite{Bsmumu,Smaria}.
\end{itemize} 
An independent good reason to pay attention to these anomalies is that the experimental error on the key observables is expected to be reduced by a large factor  in the foreseen programme in the next decade or so~\cite{LHCb,Kou:2018nap}. 

In view of all this, interesting questions to ask are: 1. Which interaction at $\Lambda_3$ can be responsible for these anomalies, assumed to persist? 2. Can this interaction be in some way related to Higgs Compositeness, which, due to the hierarchy problem, motivates a new scale in the MultiTeV?

 \begin{figure}[t]
\centering
\includegraphics[clip,width=.42\textwidth]{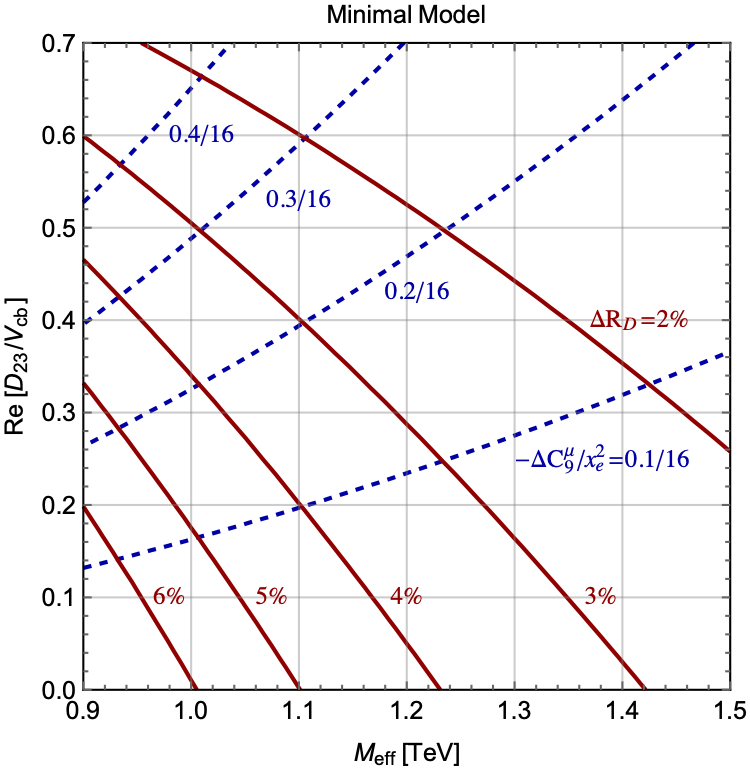}
\caption{Isolines of the charged current anomaly ($\Delta R_D$, red solid lines) and of the neutral current anomaly ($\Delta C^\mu_9= 2.1 \Delta R_K, x_e=|E^L_{23}/V_{cb}|$, blue dashed lines) from the sole interaction at $\Lambda_3$, see text. Taken from Ref.~\cite{Barbieri:2019zdz}.
}
\label{fig:MM}
\end{figure}

To the first question a candidate interaction is 
\begin{equation}
\mathcal{L}_U=g_U U_\mu^a (\bar{q}_3^a\gamma_\mu l_3)
\label{LQL}
\end{equation}
where $U_\mu^a$ is a vector leptoquark  of mass $M_U$, transforming as $(\bold{3},\bold{1})_{2/3}$ under the SM gauge group\cite{Alonso:2015sja,Calibbi:2015kma,Barbieri:2015yvd}. After going to the mass basis, as in the previous Section, this interaction is indeed capable of producing both the charged  and the neutral current anomaly.
In the parameter space $(M_{eff}=M_U/g_U, D^L_{23}, E^L_{23})$ the size of the observables $R_{D^{(*)}}$ and $R_{K^{(*)}}$ is shown in Fig.~\ref{fig:MM}, taken from Ref.~\cite{Barbieri:2019zdz}.
This is within reach of the expected future experimental precision~\cite{LHCb,Kou:2018nap}. Especially $R_{D^{(*)}}$, however, is  below the current preferred value. To get to this value a contribution from the interactions of the leptoquark at the scale $\Lambda_2$ is needed, as illustrated, e.g., in Ref.~\cite{Cornella:2019hct,Barbieri:2019zdz}.

The second question can help to address the unavoidable issue of UV-completing the vector leptoquark interaction in eq.~(\ref{LQL}). In Higgs Compositeness it is commonly assumed that the new strong interaction has a global symmetry, within which   the SM vector interactions are "weakly gauged".
In turn, always mimicking QCD,  this leads to the existence of $\rho$-like composite vectors with a mass close to $m^*$, as defined in the previous Section. Back to the leptoquark I think that it is completely natural to view it as part of composite vectors transforming as the adjoint of an $SU(4)$ global symmetry, within which the standard strong interactions are gauged. Remember that the standard quarks and leptons are spectators with respect to the new strong interaction, but, to get their coupling to the composite Higgs boson, they have to find suitable degrees of freedom to communicate with  the composite structure itself, like, e.g., composite fermions to mix with~\cite{Barbieri:2015yvd,Barbieri:2016las,Barbieri:2017tuq,Blanke:2018sro}\footnote{An apparently alternative viewpoint consists in embedding the leptoquark in a gauged $SU(4)$ group  commuting with $SU(3)\times SU(2)\times U(1)$, and with a suitable breaking pattern to generate the standard gauge interactions at low energies~\cite{Diaz:2017lit,DiLuzio:2017vat,Assad:2017iib,Calibbi:2017qbu,Bordone:2017bld}.}. 

The global $SU(4)$  symmetry, with  the same algebraic structure as the familiar Pati-Salam $SU(4)$ group~\cite{Pati:1974yy},  cannot be universally coupled to the three different families since in this case, to cope with $BR(K_L\rightarrow \mu e)< 4.7\cdot 10^{-12}$,  $SU(4)$ would have to be broken down to $SU(3)$ at a much higher scale than $\Lambda_3\approx m^*$ in the MultiTeV. Here again the appearance of more than one scale distinguishing the different families may come into play. 

The number of constraints on this interpretation of the leptoquark in flavour or even in ElectroWeak precision physics is definitely challenging. Constraints appear as well in high-$p_T$ physics, like in leptoquark pair production or, even more so, in $\tau \bar{\tau}$ production~\cite{Faroughy:2016osc,Baker:2019sli}. To see if these constraints can all be met, a useful step may be provided by the  model proposed in Ref.~\cite{Bordone:2017bld}, in particular its interpretation in 5D~\cite{Fuentes-Martin:2020pww},  which incorporates the three different scales for the different families, or a suitable modification of it. Even within obvious uncertainties, the interpretation of the anomalies as a first sign of Higgs compositeness, with flavour somehow implemented as  described in Section~\ref{AFP}, appears consistent with all observations today provided $D_{23}\approx 0$ (See Fig.~\ref{fig:Higgs_comp}) and $M_U \gtrsim 3\div 4~TeV, g_U\gtrsim 2\div 3$.

\section{Supersymmetry}

To address the hierarchy problem, supersymmetry in the MultiTeV remains the prominent if not the only weakly coupled alternative to Higgs Compositeness. In this case, however,  there are reasons to think that precision in flavour physics may not be as effective in searching for indirect signals as in the case of a strongly interacting theory, even if one blocks the possibility of highly degenerate superpartners in family space as in the case of gauge-mediated supersymmetry breaking. 

 \begin{figure}[t]
\centering
\includegraphics[clip,width=.82\textwidth]{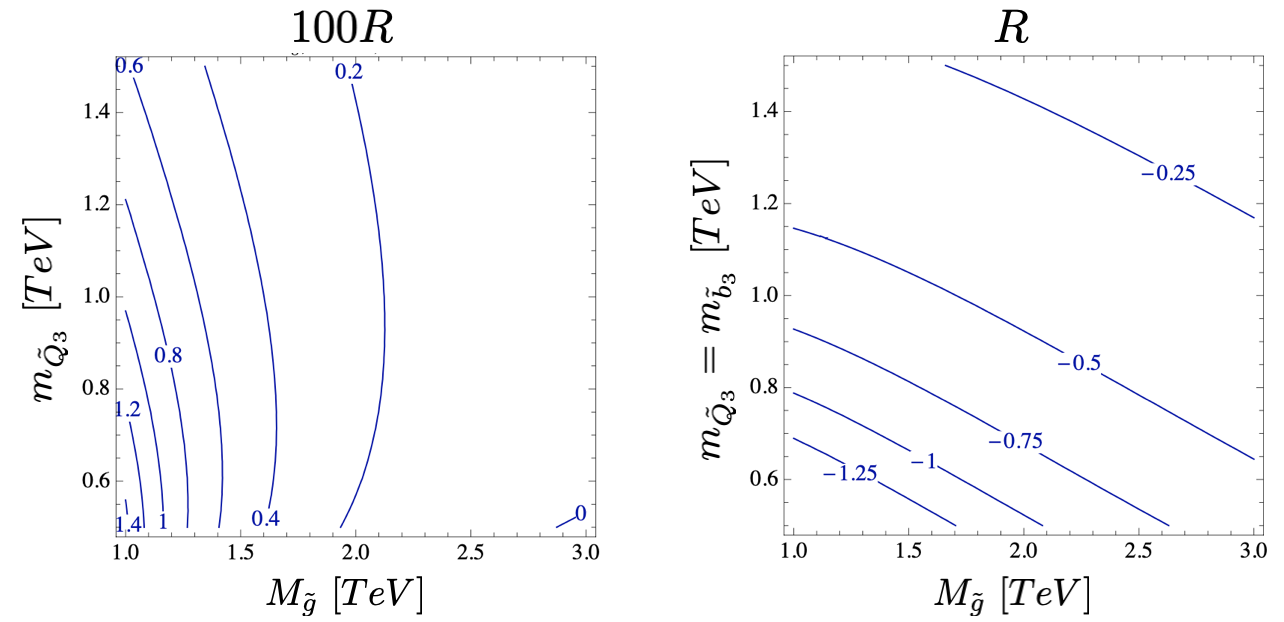}
\caption{Left:  relative correction to $\epsilon_K$, enhanced by a factor 100, from the gluino exchange box-diagram with left-handed sbottom  only. Right: 
relative correction to $\epsilon_K$  from the gluino exchange  box-diagram with left-handed and right-handed sbottoms  and  $D^R_{23}=V_{cb}$.
}
\label{fig:SUSY}
\end{figure}

Suppose that one does not assume any flavour structure at all, as in the case that leads to the bounds in Fig.~\ref{fig:DeltaF=2}. Even if these bounds are rescaled down by a factor $(g^2/4\pi)$, where $g$ stands for any of the SM gauge couplings, as necessary from a gaugino-mediated $\Delta F=2$ box diagram, many of them are left well above the MultiTeV. On the contrary, by assuming "minimally broken $U(2)^3"$, the same rescaling, this time applied to the bounds in Table~\ref{Tab:CH}, would lead to scales, i.e. the superpartner masses, normally well inside the region already explored at LHC. This is confirmed by Fig.~\ref{fig:SUSY} left, taken from Ref.~\cite{Barbieri:2014tja}, which shows the relative deviation from the SM, enhanced by a factor $100$,  in the $\Delta S=2$ parameter $\epsilon_K$ 
\begin{equation}
R\equiv \frac{\epsilon_K}{\epsilon_K^{SM}} -1,
\end{equation}
coming from the gluino exchange box diagrams, as function of the gluino and the left-handed sbottom masses, $M_{\tilde{g}}$ and 
$m_{\tilde{Q}_3}$. The current bound is $R\lesssim 0.2$.

As recalled in Section~\ref{SCH}, a key feature here is the suppression of the LR 4-fermion $\Delta S=2$ interaction, obtained by demanding  invariance under $q_3\rightarrow -q_3$ at $\Lambda_2$ or $D^R_{23}=0$ (See Section~\ref{RU}). If this requirement is abandoned, the same gluino box diagrams with $D^R_{23}=V_{cb}$ and with both left and right sbottom exchanges, taken degenerate, contribute to $R$ approximately as in Fig.~\ref{fig:SUSY} right,  adapted from Ref.~\cite{Barbieri:2014tja}. Note that $R$
 scales as $R\approx (D^R_{23}/V_{cb})^2$, thus leaving room for a possible effect in $\epsilon_K$ even for superpartner masses exceeding the current LHC bounds.

\section{Conclusions}

The  3 Scale Flavour Picture (3SFP) presented in Section~\ref{AFP}  provides an alternative to the view that the scale of flavour is far beyond the MultiTeV, as such decoupled from the physics that could explain the hierarchy problem. I admit that the construction of an actual precise theory incorporating this 3SFP is not a trivial task, hence the use of the word "picture". Nevertheless Higgs Compositeness in the MultiTeV or the confirmation of LFV in B-decays seem to require some version of the 3SFP. Furthermore, to motivate the relatively low scale required to explain the current LFV signals, if confirmed, their coupling to Higgs Compositeness, though not necessary,  is nevertheless highly suggestive, perhaps along the lines outlined in Section~\ref{AiBd} with a key role of the $SU(4)$ symmetry. The strongly reassuring fact is that  in any case precision flavour measurements in the next 10-15 years can decide on these issues~\cite{LHCb,Kou:2018nap}. In the positive case for the 3SFP, the number of different observables that would show a deviation from the SM should offer enough tools for the construction of a fully complete theory of flavour.

The role of flavour precision tests in a weakly coupled theory, like supersymmetry, appears relatively less bright, with signals that could emerge, under suitable conditions, in $\epsilon_K$ or in the electron EDM.

%
%

%
%

\section*{Acknowledgments}

I am  indebted to Gino Isidori for his collaborations on this subject and for the many discussions on most the issues of relevance to this paper.

\section*{References}
\begingroup
\renewcommand{\addcontentsline}[3]{}
\renewcommand{\section}[2]{}

\endgroup

\end{document}